\documentclass{elsart3-1-accepted}

\usepackage{amssymb}

\usepackage[english,francais]{babel}

\usepackage[applemac]{inputenc}
\usepackage{accents}

\newtheorem{theorem}{Theorem}[section]

\newtheorem{e-proposition}[theorem]{Proposition}

\newtheorem{e-definition}[theorem]{Definition\rm}

\renewcommand{\theequation}{\arabic{equation}}
\newcommand \auth {\textsc}

\newcommand \la \langle
\newcommand \ra \rangle

\newcommand \Trace   {\textrm{Tr}} 
 
\newcommand \del	    \partial

\newcommand \Mcal 	{\mathcal M}

\newcommand \Lcal 	{\mathcal L}

\newcommand \RR 		{\mathbb R}   
\newcommand \eps 	\epsilon  
\newcommand \be 		{\begin{equation}}
\newcommand \ee 		{\end{equation}} 

\let\oldmarginpar\marginpar
\renewcommand\marginpar[1]{\-\oldmarginpar[\raggedleft\footnotesize #1]%
{\raggedright\footnotesize #1}}

\def\og{\leavevmode\raise.3ex\hbox{$\scriptscriptstyle\langle\!\langle$~}}
\def\fg{\leavevmode\raise.3ex\hbox{~$\!\scriptscriptstyle\,\rangle\!\rangle$}}

\journal{Notes Comptes Rendus Acad. Sc. Paris (2010).} 
\begin{document}
\centerline{}
\begin{frontmatter}


\selectlanguage{english}
\title{Global geometry of $T^2$--symmetric spacetimes 
\\
with weak regularity}

\selectlanguage{english}
\author[authorlabel1]{Philippe G. LeFloch} and 
\ead{pgLeFloch@gmail.com}
\author[authorlabel2]{Jacques Smulevici}
\ead{jacques.smulevici@aei.mpg.de}

\address[authorlabel1]{Laboratoire Jacques-Louis Lions \& Centre National de la Recherche Scientifique, 
Universit\'e Pierre et Marie Curie (Paris 6), 4 Place Jussieu, 75252 Paris, France. Blog: philippelefloch.wordpress.com}
\address[authorlabel2]{Max-Planck-Institut f\"ur Gravitationsphysik, 
Albert-Einstein-Institut, Am M\"uhlenberg 1, 14476 Potsdam, Germany.}

\medskip
\begin{center}
{\small Received June 15, 2010. Accepted on September 9, 2010.\\
Presented by Yvonne Choquet-Bruhat}
\end{center}

\begin{abstract}
\selectlanguage{english}

We define the class of weakly regular spacetimes with $T^2$--symmetry, 
and investigate their global geometrical structure. We formulate 
the initial value problem for the Einstein vacuum equations with weak regularity, and 
establish the existence 
of a global foliation by the level sets of the area $R$ of the orbits of symmetry, 
so that each leaf can be regarded as an initial hypersurface.
Except for the flat Kasner spacetimes which are known explicitly, $R$ takes all positive values. 
Our weak regularity assumptions only require that the gradient of $R$ is continuous
while the metric coefficients belong to the Sobolev space $H^1$ (or have even less regularity).


\vskip 0.5\baselineskip

\selectlanguage{francais}
\noindent{\bf R\'esum\'e} \vskip 0.5\baselineskip \noindent
{\bf G\'eom\'etrie globale des espaces-temps $T^2$--sym\'etriques de faible r\'egularit\'e. }

Nous définissons la classe des espaces-temps à symétrie $T^2$ de faible r\'egularit\'e,  
et nous étudions leur géométrie globale. Nous formulons le problème de données initiales pour les équations
d'Einstein sous une faible régularité. 
Nous établissons l'existence d'un feuilletage global par les surfaces de niveau de la fonction d'aire $R$ 
des surfaces de symmétrie, de sorte que chaque feuille induit une hypersurface initiale. 
A l'exception des espaces-temps plats de Kasner (connus explicitement), 
la fonction $R$ prend toutes valeurs positives. 
Nos hypothèses imposent seulement que le gradient de $R$ est continu 
et que les coefficients métriques sont dans l'espace de Sobolev $H^1$ (ou sont moins réguliers). 


\end{abstract}
\end{frontmatter}



\selectlanguage{english}

\section{Introduction}
\label{intro}

We are interested in spacetimes satisfying the Einstein vacuum equations and 
we provide a description of the {\sl global geometry} of the development 
of an initial data set, having only {\sl weak regularity} but enjoying certain symmetry properties. 
We consider a manifold that is diffeomorphic to the torus $T^3$ 
and initial data that are invariant under an action of the Lie group $T^2$. 
This requirement characterizes the so-called {\sl $T^2$--symmetric spacetimes on $T^3$}.  
A large literature is available on such spacetimes when sufficiently high regularity is imposed; 
cf.~Moncrief \cite{Moncrief}, Chru\'sciel \cite{Chrusciel}, 
Berger, Chru\'sciel, Isenberg, and Moncrief \cite{BergerChruscielIsenbergMoncrief}, 
and Isenberg and Weaver \cite{IsenbergWeaver} and, for further references, \cite{Smulevici2}.  

The present Note is motivated by recent work by 
LeFloch and Stewart \cite{LeFlochStewart,LeFlochStewart2} and LeFloch and Rendall \cite{LeFlochRendall}, 
who treated Gowdy--symmetric matter spacetimes described by the Einstein-Euler equations. 
It was recognized that, due to the formation of shock waves in the fluid and by virtue of the 
Einstein equations, only weak regularity on the geometry should be allowed.

Here, we carry out this strategy further, 
and encompass (vacuum) spacetimes under less restrictive regularity and symmetry conditions.
Importantly, we present a fully geometric framework within a class of weakly regular spacetimes
and precisely identify the weak regularity conditions allowing us to, both, establish the existence of such spacetimes and tackle the analysis of their global geometric structure.  

Even in the smooth class, there remain challenging open problems for these spacetimes,
such as the question of geodesic completeness in the expanding direction 
or the structure of singularities in the contracting direction, 
and we believe that the new framework and estimates developed in the present work
will be useful to tackle some of these issues. 

The initial value problem in general relativity (in the vacuum case) is formulated as follows. 
An initial data set for the vacuum Einstein equations is a triple $(\Sigma,h,K)$ such that 
$(\Sigma,h)$ is a Riemannian $3$-manifold and 
$K$ is a symmetric $2$-tensor field on $\Sigma$, 
satisfying the so-called Einstein's constraint equations.   
A solution to the initial value problem  with initial data $(\Sigma,h,K)$ 
is a $(3+1)$-dimensional Lorentzian manifold $(\Mcal,g)$ satisfying the vacuum Einstein equations
(that is, 
the Ricci flat condition $\mbox{Ric}=0$), 
together with an embedding $\phi: \Sigma \to \Mcal$ such that $\phi(\Sigma)$ is a Cauchy hypersurface of $(\Mcal,g)$ 
and the pull-back of its first and second fundamental forms coincides with $h$ and $K$, respectively. 

Recall that the existence of a unique (up to diffeomorphism) maximal globally hyperbolic 
development $(\Mcal,g)$ was established in pioneering work by 
Choquet-Bruhat and Geroch \cite{ChoquetBruhat1952,ChoquetBruhatGeroch}
(reviewed in \cite{ChoquetBruhat2009} together with extensive recent material). 
The optimal existence result as far as the regularity of the initial data is concerned
is currently given in Klainerman and Rodnianski \cite{KlainermanRodnianski} in the Sobolev space $H^{2+\eps}$ (with $\eps>0$).  


\section{Statement of the main result}

By restricting attention to the class of $T^2$--symmetric spacetimes, 
our regularity assumptions will go far below the regularity 
covered in previous works.  
We assume here that the area $R$ of the orbits of $T^2$--symmetry is a $C^1$ function. 
Introducing a frame of commuting vector fields $(T,\Theta,X,Y)$ where $X,Y$ are Killing fields,  
the remaining components of the metric belong to the Sobolev space $H^1$ 
on every spacelike slice of a foliation, 
except for the cross terms between the Killing fields $X,Y$ and the vector field $\Theta$
which have lower regularity.  (See Section~3, below.)  
Importantly, {\sl not all} Christoffel symbols and curvature components make sense as distributions, but 
only those that are relevant to formulate the Einstein equations in a weak sense. 
To do otherwise, additional regularity would be necessary, as recognized in \cite{LeFlochMardare}.

Under these regularity conditions, our main result 
establishes the existence of a weakly regular development,
and provides detailed information about its global geometry.  
We refer to the next section and \cite{LS} for the precise terminology used in the following theorem.

\begin{theorem}
\label{theorem61}

\begin{itemize}

\item If $(\Sigma, h, K)$ is a weakly regular $T^2$--symmetric triple, 
then the Einstein constraint equations (suitably reformulated) make sense as equality between distributions. 
Similarly, 
if $(\Mcal,g)$ is a weakly regular $T^2$--symmetric Lorentzian manifold,
then the Einstein equations (suitably reformulated) make sense as equality between distributions.  

\item For all weakly regular $T^2$--symmetric initial data set $(\Sigma,h,K)$
with topology $T^3$ and constant area function $R$, 
there exists a weakly regular vacuum spacetime with $T^2$--symmetry on $T^3$, $(\Mcal,g)$, which is a development of $(\Sigma, h, K)$ and admits a global foliation by the level sets of the area 
$R \in (R_*,\infty)$ (for some $R_* \geq 0$). 
Moreover, unless $(\Mcal,g)$ is a flat Kasner spacetime, one has $R_* = 0$.

\end{itemize}

\end{theorem}

The compactness arguments developed in \cite{LS} for the existence proof  
are very different from the classical techniques used in the high regularity 
case~\cite{Moncrief,BergerChruscielIsenbergMoncrief}. In the rest of this Note, we discuss in further detail the class of initial data and present our new
geometric formulation which is an essential contribution of \cite{LS}.


\section{Weakly regular $T^2$--symmetric manifolds}

While all topological manifolds under consideration are of class $C^\infty$,   
the metric structures of interest here have low regularity. 
For instance, the Lie derivative $\Lcal_Z h$ of a measurable and locally integrable $2$-tensor 
$h$ is defined in the distribution sense, for any $C^1$ vector fields $X,T,Z$, by  
\be
\label{284}
(\Lcal_X h) (T,Z) := X(h(T,Z)) - h(\Lcal_X T,Z) - h(T, \Lcal_X Z),  
\ee
in which the last two terms are classically defined as locally integrable functions,
but the first one is defined in the distributional sense, only.

\begin{e-definition} 
\label{def21}
A {\bf weakly regular $T^2$--symmetric Riemannian manifold} is a compact $C^\infty$ manifold $\Sigma$ 
endowed with a tensor field $h$ which enjoys the following regularity and symmetry properties:
\begin{enumerate}

\item {\bf Riemannian structure.} The field $h$ is a Riemannian metric with components in $L^\infty(\Sigma)$,
and the associated volume form is continuous.  

\item {\bf Symmetry property.} On the Riemannian manifold $(\Sigma,h)$, there exists an action of the Lie group $T^2$ 
generated 
by two (linearly independent, commuting) Killing fields $X,Y$ of class $C^\infty$ (with closed orbits), 
satisfying in particular 
$\Lcal_X h =0$ and $\Lcal_Y h = 0$  
understood in the distributional sense (\ref{284}). 

\item {\bf Regularity of the metric on the orbits.} The functions $h(X,X)$, $h(X,Y)$, and $h(Y,Y)$ 
belong to the Sobolev space $H^{1}(\Sigma;h)$.

\item {\bf Regularity of the area function.} 
The orbit area function 
\be
\label{ass:rconstant}
R^2 := h(X,X) \, h(Y,Y) - h(X,Y)^2
\ee  
(which then lies in $W^{1,1}(\Sigma;h)$) is required to belong to $C^1(\Sigma)$.  

\item {\bf Regularity of the metric on the orthogonal of the orbits.}
There exists a {\rm smooth} vector field $\Theta$ defined on $\Sigma$ such that $(\Theta,X,Y)$ form a basis of commuting vector fields and  the vector field  
\be
\label{395}
Z := \Theta + a \, X + b \, Y, \qquad Z \in \big\{ X,Y \big\}^\perp, \qquad \Lcal_X Z = \Lcal_Y Z = 0 
\ee
implies the regularity $h(Z,Z) \in H^1(\Sigma;h)$. 

\end{enumerate} 
\end{e-definition}

We refer to such a triple $(X,Y,Z)$ as an {\sl adapted frame} on $\Sigma$. We emphasize that no regularity is required on the derivatives of the ``cross-terms'' $h(X,\Theta)$ and $h(Y,\Theta)$. 
We also need to consider the second fundamental form of the initial slice.

\begin{e-definition} 
\label{def:t2tri}
A {\bf weakly regular $T^2$--symmetric triple} $(\Sigma,h,K)$
is a weakly regular $T^2$--symmetric Riemannian manifold $(\Sigma,h)$ with adapted frame $(X,Y,Z)$,  
satisfying the following conditions: 
\begin{enumerate}

\item {\bf Integrability.} 
The field $K$ is a symmetric $2$-tensor field on $\Sigma$ such that its components in the basis $(X,Y,Z)$ lies in $L^2(\Sigma)$. 

\item {\bf Symmetry property.} 
Moreover, $K$ is invariant under the action of the Lie group $T^2$ generated by $(X,Y)$, that is, 
$\Lcal_X K = \Lcal_Y K = 0$ 
understood in the sense of distributions (\ref{284}). 

\item {\bf Additional regularity} (associated with the time derivative of the area function): 
$$
\Trace^{(2)} (K) := h(X,X) \, K(X,X) + 2 h(X,Y) \, K(X,Y) + h(Y,Y) \, K(Y,Y) \in C(\Sigma). 
$$   

\end{enumerate}
\end{e-definition}

Our notion of initial data set follows the above two definitions, as we define the curvature of the manifold and impose the constraint equations in the distribution sense.  
Such a statement requires to take into account our symmetry assumption in order for the curvature terms to 
make sense as distributions.

\section*{\vskip-.9cm Acknowledgments}

\vskip-.4cm The first author (PLF) was partially supported by
the Agence Nationale de la Recherche (ANR) through the grant 06-2-134423 entitled
``Mathematical Methods in General Relativity'' (Math-GR). 
The second author (JS) thanks the Albert Einstein Institute for financial support. 
 This work was done when the second author (JS) was visiting the Laboratoire Jacques-Louis Lions 
 with
 the support of the ANR. 


\end{document}